\input{aipcheck}

\documentclass[
    ,final            
  ]
  {aipproc}

\layoutstyle{6x9}

\begin{document}

\title{Flat direction MSSM (A-term) Inflation}


\author{Juan Garc\'\i a-Bellido}{
  address={Dep. F\'\i sica Te\'orica, C-XI, Univ.
Aut\'onoma de Madrid, Cantoblanco, 28049 Madrid, Spain}
}

\begin{abstract}
The Minimal Supersymmetry Standard Model contains several hundreds
of D- and F-flat directions that are lifted by soft susy breaking
terms as well as non-renormalizable terms. In a recent paper we
find that only two of these directions, {\bf LLe} and {\bf udd}
can accommodate inflation. The model predicts more than $10^3$
e-foldings with an inflationary scale of $H_{\rm inf}\sim 1$ GeV,
provides a tilted spectrum compatible with WMAP3, and a negligible
tensor perturbation. The reheating temperature could be as low as
$T_{rh}\sim 1$~TeV. The model is stable under radiative as well as
supergravity corrections, although a significant finetunning has
to be imposed on the ratio of two parameters of the model. The
RGE equations allow us to relate the model parameters with slepton
and gaugino masses explorable in the LHC, while the neutralino in
this model could well be within reach of present dark matter searches
in the laboratory.
\end{abstract}

\maketitle


\section{Introduction}

Decades of research have gone by with the purpose of finding a
connection between the early universe and the fundamental interactions
of particle physics.  There seems to be consensus nowadays within the
community that the early universe is best described by a short period
of inflation, which sets up the initial conditions for the subsequent
hot Big Bang. In the absence of a theory of inflation, cosmologists
have proposed a plethora of different models more or less based of
fundamental theories of particle interactions. In spite of the
tremendous growth in data provided by the recent revolution in
cosmological observations, from the cosmic microwave background to the
distribution of matter of large scales, we still do not know much
about the true model of inflation. In particular, we don't know what
was the fundamental scale at which it took place, whether at GUT
scales (of order $10^{16}$ GeV) or at EW scales (of order 100 GeV),
while both are still compaticle with present observations. On the
other hand, particle physicists have searched for decades for
consistent theories of (mostly supersymmetric) extensions of the
Standard Model of particle interactions. With the imminent advent of
the Large Hadron Collider at CERN we will finally explore the symmetry
breaking sector of the electroweak theory and perhaps discover new
particles associated with the longsoughtafter extensions. It is thus
natural to search for theories of inflation based on the low energy
supersymmetric Standard Model. These theories typically have too large
masses and couplings to sustain inflation, except along certain well
defined directions in the scalar potential that are D- and
F-flat. However, not all of these directions remain flat; most of them
get lifted by radiative and supergravity corrections, as well as by
soft-susy breaking and higher order terms in the superpotential.

In a recent paper~\cite{AEGM}, we have proposed a model of inflation
based on the {\bf udd} and {\bf LLe} flat directions of Minimally
Supersymmetric Standard Model (for a review of MSSM flat directions,
see~\cite{GKM,KARI-REV}). In this model the inflaton is a gauge
invariant combination of either squark or slepton fields.  For a
choice of the soft SUSY breaking parameters $A$ and the inflaton mass
$m_\phi$, the potential along the flat {\bf udd} and {\bf LLe}
directions is such that there is a period of slow roll inflation of
sufficient duration to provide the observed spectrum of CMB
perturbations and an unambiguous prediction of the spectral tilt. In
the inflationary part of the MSSM potential the second derivative is
vanishing and the slow roll phase is driven by the third derivative of
the potential.

MSSM inflation occurs at a very low scale with $H_{\rm inf}\sim 1-10$~GeV
and with field values much below the Planck scale $\phi_0 \sim 10^{14} -
10^{15}$ GeV. Hence it stands in
strong contrast to the conventional inflation models which are based
on ad hoc gauge singlet fields and often employ field values close to
Planck scale (for a review, see ~\cite{LYTH}). In such models the
inflaton couplings to SM physics are unknown.  As a consequence much
of the post-inflationary evolution, such as reheating, thermalization,
generation of baryon asymmetry and cold dark matter, is not calculable from
first principles. The great virtue of MSSM inflation based on flat
directions is that the inflaton couplings to Standard Model particles
are known and, at least in principle, measurable in laboratory
experiments such as LHC or a future Linear Collider.

However, as in almost all inflationary models, a fine tuning of the
initial condition is needed to place the flat direction field $\phi$
to the immediate vicinity of the saddle point $\phi_0$ at the onset of
inflation. In addition, there is the question of the stability of the
saddle point solution and of the existence of a slow roll regime.
These are issues that we wish to address in detail in the present
paper. Both supergravity and radiative corrections to the flat
direction inflaton potential must be considered.  Hence we need to
write down and solve the renormalization group (RG) equations for the
MSSM flat directions of interest. RG equations are also needed to
scale the model parameters, such as the inflaton mass, down to TeV
scale; since the inflaton mass is related either to squark or slepton
masses, it could be measured by LHC or a future Linear Collider.

Because the inflaton couplings to ordinary matter are known, inflaton
decay and thermalization are processes that can be computed in an
unambiguous way. Unlike in many models with a singlet inflaton, in MSSM
inflation the potential relevant for decay and thermalization cannot
be adjusted independently of the slow roll part of the potential.


\section{The Model}

Let us begin by considering a flat direction ${\phi}$ with a
non-renormalizable superpotential term,
$W = (\lambda_n/n)(\Phi^n/M^{n-3}_{\rm P})$,
where $\Phi$ is the superfield which contains the flat direction.
Within MSSM all the flat directions are lifted by $n=9$
non-renormalizable operator~\cite{GKM}.  Together with the
corresponding $A$-term and the soft mass term, it gives rise to the
following scalar potential for ${\phi}$:
\begin{equation} \label{scpot}
V = {1\over2} m^2_\phi\,\phi^2 + A\cos(n \theta  + \theta_A)
{\lambda_{n}\phi^n \over n\,M^{n-3}_{\rm P}} + \lambda^2_n
{{\phi}^{2(n-1)} \over M^{2(n-3)}_{\rm P}}\,,
\end{equation}
where $m_\phi$ is the soft SUSY breaking mass for $\phi$. Here $\phi$
and $\theta$ denote the radial and the angular coordinates of the
complex scalar field $\Phi=\phi\,\exp[i\theta]$ respectively, while
$\theta_A$ is the phase of the $A$-term (thus $A$ is a positive
quantity with a dimension of mass). Note that the first and third
terms in Eq.~(\ref{scpot}) are positive definite, while the $A$-term
leads to a negative contribution along the directions where $\cos(n
\theta + \theta_A) < 0$.

The maximum impact from the $A$-term is obtained when $\cos(n \theta +
\theta_A) = -1$ (which occurs for $n$ values of $\theta$).  Along
these directions $V$ has a secondary minimum at $\phi = \phi_0
\sim \left(m_\phi M^{n-3}_{\rm P}\right)^{1/n-2} \ll M_{\rm P}$
(the global minimum is at $\phi=0$), provided that
\begin{equation}\label{cond}
A^2 \geq 8 (n-1) m^2_{\phi}\,.
\end{equation}
At this minimum the curvature of the potential along the radial
direction is $\,+ m^2_{\phi}$ (it is easy to see that the curvature is
positive along the angular direction, too), and the potential reduces
to: $V \sim m_\phi^2\phi_0^2 \sim m_{\phi}^2\left(m_{\phi} M^{n-3}_{\rm
P}\right)^{2/(n-2)}$.  Now consider the situation where the flat
direction is trapped in the false minimum $\phi_0$. If its potential
energy, $V$, dominates the total energy density of the Universe, a
period of inflation is obtained. The Hubble expansion rate during
inflation will then be
$H_{\rm inf} \sim (m_\phi \phi_0 / M_{\rm P}) \sim
m_\phi (m_\phi /M_{\rm P})^{1/(n-2)}$.
Note that $H_{\rm inf} \ll m_\phi$, which implies that the potential is
too steep at the false minimum and $\phi$ cannot climb over the
barrier which separates the two minima just with the help of quantum
fluctuations during inflation.

The potential
barrier disappears when the inequality in Eq.~(\ref{cond}) is
saturated, i.e. when $A^2 = 8 (n-1) m^2_\phi$.
Then both the first and second derivatives of $V$ vanish at $\phi_0$,
i.e. $V^{\prime}(\phi_0)=0,~ V^{\prime\prime}(\phi_0)=0$, and the
potential becomes very flat along the {\it real direction}. Around
$\phi_0$ the field is stuck in a plateau with potential energy
\begin{equation}\label{phi0}
V(\phi_0) = {(n-2)^2\over2n(n-1)}\,m^2_\phi \phi_0^2\,, \hspace{5mm}
\phi_0 = \left({m_\phi M^{n-3}_{\rm P}\over
\lambda_n\sqrt{2n-2}}\right)^{1/(n-2)}\,.
\end{equation}
However, although the second derivative of the potential vanishes, the
third does not; instead $V'''(\phi_0) = 2(n-2)^2\,m^2_\phi / \phi_0$.
Around $\phi=\phi_0$ we can thus expand the potential as $V(\phi) =
V(\phi_0) + (1/ 3!)V'''(\phi_0)(\phi-\phi_0)^3$. Hence, in the range
$[\phi_0 - \Delta \phi, \phi_0 + \Delta \phi]$, where $\Delta \phi
\sim H_{\rm inf}^2/V^{\prime\prime\prime}(\phi_0) \sim
\left({\phi}^3_0/M^2_{\rm P}\right) \gg H_{\rm inf}$, the real direction
has a flat potential.
We can now solve the equation of motion for the $\phi$ field in the
slow-roll approximation, $3H\dot\phi
=-(V'''(\phi_0)/2)(\phi-\phi_0)^2$. Note that the field only feels
the third derivative of the potential.  Thus, if the initial
conditions are such that the flat direction starts in the vicinity of
$\phi_0$ with $\dot\phi\approx 0$, then a sufficiently large number of
e-foldings of the scale factor can be generated. In fact, quantum
fluctuations along the tachyonic direction~\cite{LINDE} will drive the
field towards the minimum. However, quantum diffusion is stronger than
the classical force, $H_{\rm inf}/2\pi > \dot\phi/H_{\rm inf}$, for
$(\phi_0-\phi)/\phi_0 \leq ({m_\phi\phi_0^2 / M_{\rm P}^3})^{1/2}\,,$
but from then on, the evolution is determined by the usual slow roll.
A rough estimate of the number of e-foldings is then given by
\begin{equation} \label{efold}
N_e(\phi) = \int {H_{\rm inf} d\phi \over \dot\phi}
\simeq {\phi^3_0 \over 2n(n-1) M^2_{\rm P} (\phi_0-\phi)} ~ ,
\end{equation}
where we have assumed $V'(\phi) \sim (\phi - \phi_0)^2 V'''(\phi_0)/2$
(this is justified since $V'(\phi_0) \sim 0, V''(\phi_0)\sim 0$). Note
that the initial displacement from $\phi_0$ cannot be smaller than
$H_{\rm inf}$, due to the uncertainty from quantum fluctuations.

Inflation ends when $\epsilon\sim1$, or
\begin{equation}
{(\phi_0-\phi) \over \phi_0} \sim
\Big({\phi_0 \over 2n(n-1)M_{\rm P}}\Big)^{1/2}\,.
\end{equation}
After inflation the coherent oscillations of the flat direction excite
the MS(SM) degrees of freedom and reheat the universe.

\begin{figure}
  \includegraphics[height=8cm]{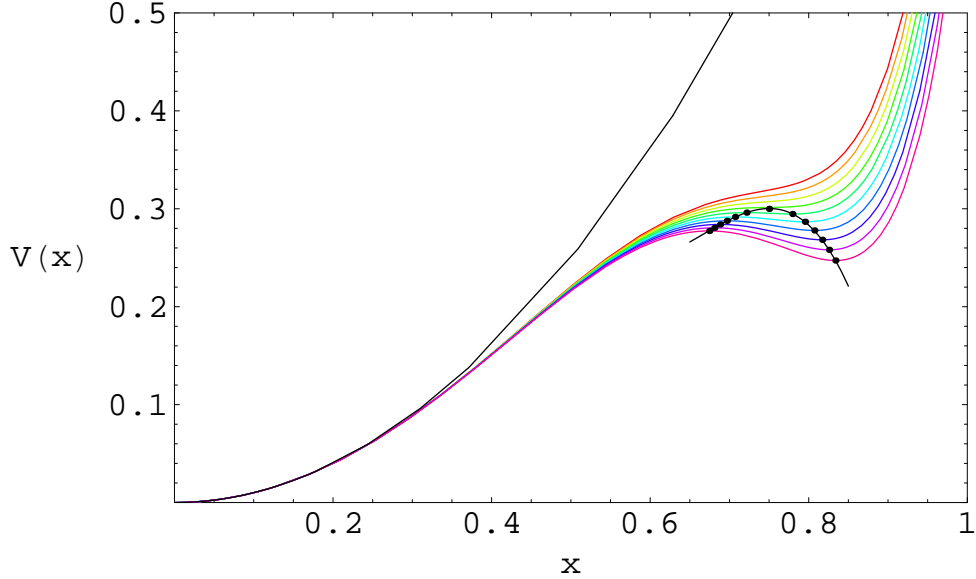}
  \caption{The colored curves depict the full potential, where
$V(x)\equiv V(\phi)/(0.5~m_{\phi}^2 M_{\rm P}^2(m_{\phi}/M_{\rm
P})^{1/2})$, and $ x\equiv (\lambda_n M_{\rm P}/m_{\phi})^{1/4}
(\phi/M_{\rm P})$. The black curve is the potential arising from the
soft SUSY breaking mass term. The black dots on the colored potentials
illustrate the gradual transition from minimum to the saddle point and
to the maximum.}
\end{figure}

Let us now identify the possible MSSM inflaton candidates. Recall
first that the highest order operators which give a non-zero $A$-term
are those with $n=6$. This happens for flat directions represented by
the gauge invariant monomials $\phi={\bf L_i L_j e_k}\,; \
\phi= {\bf u\,d_i d_j}$.
The flatness of the potential require that $i \neq j \neq k$ in the
former and $i \neq j$ in the latter. For $n=6$ and $m_{\phi} \sim 1$
TeV, as in the case of weak scale supersymmetry breaking, we find the
following generic results:\\
a) {\it Sub-Planckian VEVs}: In an effective field theory where
the Planck scale is the cut-off, inflationary potential can be trusted
only below the Planck scale, usually a challenge for the model
building. In our case the flat direction VEV is
sub-Planckian for the non-renormalizable operator $n=6$, i.e.
$\phi_0\sim 1-3\times10^{14}$~GeV for $m_{\phi}\sim 1-10$~TeV, while
the vacuum energy density ranges $V\sim 10^{34}-10^{38}~({\rm GeV})^4$
(we assume $\lambda_{n}=1$; generically $\lambda_{n}\leq 1$ but its
precise value depends on the nature of high energy physics);\\
b) {\it Low scale inflation}: Although it is extremely hard to
build an inflationary model at low scales, for the energy density
stored in the MSSM flat direction vacuum, the Hubble expansion rate
comes out as low as $H_{\rm inf}\sim 1-10$~GeV.  It might be possible
to lower the scale of inflation further to the electroweak scale;\\
c) {\it Enough e-foldings}: At low scales, $H_{\rm inf}\sim {\cal
O}(1)$~GeV, the number of e-foldings, $N_{\rm COBE}$, required
for the observationally relevant perturbations, is much less than
$60$~\cite{LEACH}. In fact the number depends on when the Universe
becomes radiation dominated (note that full thermalization is not
necessary as it is the relativistic equation of state which matters).

If the inflaton decays immediately after the end of inflation, which
has a scale $V\sim 10^{36}~({\rm GeV})^4$, we obtain $N_{\rm
COBE} \sim 47$. The relevant number of e-foldings could
be greater if the scale of inflation becomes larger. For instance, if
$m_{\phi}\sim 10$~TeV, and $V\sim 10^{38}~({\rm GeV})^4$, we have
$N_{\rm COBE}\sim 50$. For the MSSM flat direction lifted by
$n=6$ non-renormalizable operators, we obtain the total number of
e-foldings as $N_{e} \sim (\phi_0^2 / m_\phi M_{\rm P})^{1/2} \sim
10^3$, computed from the end of diffusion.
This bout of inflation is sufficiently long to produce
a patch of the Universe with no dangerous relics. Domains
initially closer to $\phi_0$ will enter self-reproduction in
eternal inflation.

Let us now consider adiabatic density perturbations. Despite the low
inflationary scale, $H_{\rm inf}\sim 1-10$ GeV, the flat
direction can generate adequate density perturbations as required to
explain the COBE normalization. This is due to the extreme flatness of
the potential (recall that $V'=0$), which causes the velocity of the
rolling flat direction to be extremely small.  Thus we find an
amplitude of
\begin{equation}
\label{amp}
\delta_{H}\simeq \frac{1}{5\pi}\frac{H^2_{inf}}{\dot\phi}
\sim {m_\phi M_{\rm P}\over \phi_{0}^2}\,N_{\rm COBE}^2 \sim 10^{-5}\,,
\end{equation}
for $m_{\phi}\sim 10^{3}-10^{4}$~GeV, where $\phi_0$ is given by
Eq.~(\ref{phi0}).  In the above expression we have used the slow roll
approximation $\dot\phi\simeq -V'''(\phi_0)(\phi_0- \phi)^2/3H_{\rm
inf}$, and Eq.~(\ref{efold}).  Note the importance of the $n=6$
operators lifting the flat directions ${\bf LLe}$ and ${\bf
udd}$. Higher order operators would have allowed for larger VEVs and a
large $\phi_0$, therefore leaving the amplitude of the perturbations
too low.

The spectral tilt of the power spectrum is not negligible because,
although $\epsilon\sim1/N_{\rm COBE}^4\ll 1$, the parameter $\eta =
-2/N_{\rm COBE}$ and thus
\begin{equation}\label{spect}
n_s = 1 + 2\eta - 6\epsilon \simeq 1 - {4\over N_{\rm COBE}} \sim 0.92\,,
\hspace{5mm}
{d\,n_s\over d\ln k} = - {4\over N_{\rm COBE}^2} \sim - 0.002\,,
\end{equation}
which agrees with the current WMAP 3-years' data within
$2\sigma$~\cite{WMAP}, while there are essentially no tensor modes.
Note that the tilt can be tuned to match the central value of the
WMAP 3-years' by chosing $\lambda_{n}\leq 1$.

Recall that quantum loops result in a logarithmic running of the soft
supersymmetry breaking parameters $m_\phi$ and $A$. One might then
worry about their impact on Eq.~(\ref{cond}) and the success of
inflation. Note however that the only requirement is that one must 
use the VEV-dependent values of $m_\phi(\phi)$ and $A(\phi)$ in
Eq.~(\ref{cond}) for determining $\phi_0$. We have checked that the
crucial ingredient for a successful inflation, i.e. having a very
flat potential such that $V'(\phi_0) = V''(\phi_0) = 0$, remains true
after quantum corrections. The only difference is a small shift in 
the value of $\phi_0$.

After the end of inflation, the flat direction eventually starts
rolling towards its global minimum. The flat direction decays into
light relativistic MS(SM) particles which obtains kinetic equilibrium
rather quickly.
Although the plasma heats up to a large value due to large momenta of
the inflaton decay products, the process of thermalization, which
requires chemical equilibrium, can be a slow process.
Just to illustrate, we note that within MSSM there are other flat directions
orthogonal to the inflaton. These can develop large
VEVs and induce large masses to the MSSM quanta, i.e. squarks and
sleptons and gauge bosons and gauginos. As a consequence, there
will be a kinematical blocking for the inflaton to
decay~\cite{ROUZ-REHEAT} which can delay thermalization.
The details of thermalization would require involved calculations.
However, perhaps the
best guess is to assume that the flat direction mass gives the lower
limit on a temperature, where all the MSSM degrees of freedom are in
thermal equilibrium, $T_{rh}\sim m_{\phi}\sim {\cal O}(1-10)~{\rm TeV}$.
Note that this temperature is sufficiently high for cold electroweak
baryogenesis~\cite{BARYO-REV,CEWB} and for both thermal and non-thermal
cold dark matter production~\cite{CDM-REV,Ester}.

\section{Conclusion}

The existence of a saddle point in the scalar potential of the ${\bf
udd}$ or ${\bf LLe}$ MSSM flat directions appears, perhaps
surprisingly, to provide all the necessary ingredients for an
observationally realistic model of inflation~\cite{AEGM}. MSSM
inflation takes place at a low energy scale so that it is naturally
free of supergravity and super-Planckian effects. The exceptional
feature of the model, which sets it apart from conventional singlet
field inflation models, is the fact that here the inflaton is a gauge
invariant combination of the squark or slepton fields. As a
consequence, the couplings of the inflaton to the MSSM matter and
gauge fields are known.  This makes it possible to address the
questions of reheating and gravitino production in an unambiguous way.
Since ${\bf udd}$ and ${\bf LLe}$ are
independently flat, therefore, if ${\bf LLe}$ is the inflaton, the
${\bf udd}$ direction can also acquire a large VEV
simultaneously. This gives a large mass to gluons/gluinos which
decouples them from the thermal bath, and hence suppresses thermal
gravitino production. Non-thermal production
of gravitinos is negligible in our model.

In the MSSM inflation model the mass of the inflaton is not a free
parameter but is related to the masses of e.g. sleptons, should the
${\bf LLe}$ direction be the inflaton. We have solved the appropriate
RG equation equations to relate the inflaton mass to the slepton masses at
energies accessible to accelerators such as LHC and found that LHC can
indeed put a constraint on the model: it may not be able to verify it,
but it certainly can rule it out.

The model predictions are not modified by supergravity corrections,
i.e. the observables are insensitive to the nature of Kh\"aler
potential. MSSM inflation also illustrates that it is free from any
Trans-Planckian corrections.  MSSM inflation retains the successes of
thermal production of LSP as a dark matter and the electroweak
baryogenesis within MSSM.

The existence of the saddle point requires fine-tuning~\cite{SDL}.  As
shown in Ref.~\cite{Aterm}, the parameter $\alpha$ should be small.
In that paper we dealt both with the local minimum (and the
constraints imposed by efficient tunneling from the local minimum onto
the slow roll part of the potential) as well as with the inflection
point, $V'(\phi)>0$.  We found that a fine-tuning of the order of one
part in $10^9$ is sufficient. We also found that one-loop radiative
corrections induce a shift in the ratio $\alpha$ which is of the order
of $10^{-2}$, making it necessary to adjust the ratio order by order
up to four loops in the perturbative expansion. However, it is
conceivable that the mechanism of supersymmetry breaking could remove
the fine-tuning in some natural, dynamical way. For instance, $A/m$
could turn out to be a renormalization group invariant so that once
the ratio is fixed e.g. by threshold corrections, it would remain
fixed at all orders. This is just speculation, of course, but perhaps
warranted by the simplicity and the apparent success of MSSM flat
direction inflation, which is unique in being both a successful model
of inflation and at the same time having a concrete and real
connection to physics that can be observed in earth bound
laboratories.


\begin{theacknowledgments}

I thank my collaborators, Rouzbeh Allahverdi, Kari Enqvist, Asko
Jokinen, and very specially Anupam Mazumdar, for their entertaining
and often passionate discussion of physics. I also thank the
organisers of the workshop ``The Dark Side of the Universe'' for
a very enjoyable meeting. My research has been supported in part by 
a MEC-CICYT project FPA2003-04597, as well as by the European Union
Training Network MRTN-CT-2006-035863. I would also like to thank
the Galileo Galilei Institute for Theoretical Physics for the
hospitality during my stay in Florence, and the INFN for partial
support during the completion of this work. 

\end{theacknowledgments}


\end{document}